\documentstyle[12pt,epsfig]{article}
\newcommand{\be}{\begin{equation}}
\newcommand{\ee}{\end{equation}}
\def\aprle{\buildrel < \over {_{\sim}}}

\begin{document}
\topmargin 0pt
\oddsidemargin=-0.4truecm
\evensidemargin=-0.4truecm
\renewcommand{\thefootnote}{\fnsymbol{footnote}}
\newpage
\setcounter{page}{1}
\begin{titlepage}     
\vspace*{-2.0cm}
%%%\vspace*{-1.0cm}
\begin{flushright}
FISIST/18-99/CFIF \\
%%\vspace*{-0.2cm}
hep-ph/9911364
\end{flushright}
\vspace*{0.5cm}
\begin{center}
{\Large \bf Seesaw mechanism and structure of neutrino \\ 
\vspace{0.1cm}
mass matrix}
%%%\vspace{0.5cm}
\vspace{1.0cm}

{\large E. Kh. Akhmedov\footnote{On leave from National Research Centre 
Kurchatov Institute, Moscow 123182, Russia. 
E-mail: akhmedov@gtae2.ist.utl.pt}, 
G. C. Branco\footnote{E-mail: d2003@beta.ist.utl.pt} 
and M. N. Rebelo\footnote{E-mail: rebelo@beta.ist.utl.pt} }\\
\vspace{0.05cm}
{\em Centro de F\'\i sica das Interac\c c\~oes Fundamentais (CFIF)} \\
{\em Departamento de F\'\i sica, Instituto Superior T\'ecnico }\\
{\em Av. Rovisco Pais, P-1049-001, Lisboa, Portugal }\\
\end{center}
\vglue 0.8truecm
\begin{abstract}
We consider the seesaw mechanism of neutrino mass generation in 
the light of our present knowledge of the neutrino masses and 
mixing. We analyse the seesaw mechanism constrained by the following
assumptions: (1) minimal seesaw with no Higgs triplets, (2) hierarchical 
Dirac masses of neutrinos, (3) large lepton mixing primarily or solely due 
to the mixing in the right-handed neutrino sector, and (4) unrelated 
Dirac and Majorana sectors of neutrino masses. We show that large mixing 
governing the dominant channel of the atmospheric neutrino oscillations 
can be naturally obtained and point out that this constrained seesaw mechanism 
favours the normal mass hierarchy for the light neutrinos leading to a small 
$V_{e3}$ entry of the lepton mixing matrix and a mass scale of the lightest 
right handed neutrino $M\simeq 10^{10} - 10^{11}$ GeV. Any of the three main 
neutrino oscillation solutions to the solar neutrino problem can be 
accommodated. The inverted mass hierarchy and quasi-degeneracy of neutrinos 
are disfavoured in our scheme.  
\end{abstract}
%\vspace{1.cm}
%%\centerline{Pacs numbers: 14.60.+Pq, 26.65.+t} 
%\vspace{.5cm}
%%\centerline{Keywords: neutrinos, earth, matter effects} 
%\vspace{.5cm}
%\vspace{.3cm}
\end{titlepage}
\renewcommand{\thefootnote}{\arabic{footnote}}
\setcounter{footnote}{0}
\newpage
\section{Introduction}

The study of neutrino masses and lepton mixing may provide crucial clues 
towards the solution of the general problem of fermion masses and mixing. 
One of the striking features of lepton mixing is the fact that an explanation 
of the atmospheric neutrino anomaly \cite{ANA} through neutrino oscillations 
requires a large mixing angle for $\nu_\mu\to \nu_\tau$ or $\nu_\mu\to 
\nu_{sterile}$. This is to be contrasted with the quark sector, where the 
mixing is small. Neutrinos have a distinguishing feature of being the only 
known fermions which are neutral with respect to all conserved charges, 
namely electric charge and colour. As a result, they can have Majorana 
masses, which in particular can arise through the seesaw mechanism 
\cite{seesaw}. 

The seesaw mechanism provides a very natural and attractive explanation of 
the smallness of the neutrino masses compared to the masses of the charged 
fermions of the same generation through the existence of heavy singlet
neutrinos $\nu_R$ 
\footnote{{For recent studies of the seesaw mechanism see, e.g., 
\cite{recent,JS,AFM}.}}. 
However, this mechanism does not fix completely the overall scale of the
light neutrino masses since the mass scale of $\nu_R$, though
naturally large, is not precisely known. Moreover, the ratios of the light
neutrino masses as well as the lepton mixing angles remain arbitrary: one
can easily obtain any desired values of these parameters by properly choosing 
the neutrino Dirac mass matrix and Majorana mass matrix of singlet neutrinos. 
Therefore by itself, without any additional assumptions, this mechanism
has limited predictive power. 
To gain more predictivity one has to invoke additional assumptions. 
In this letter, we study how phenomenologically viable neutrino 
masses and mixings can be generated within the framework of the seesaw 
mechanism, together with a reasonable set of assumptions, which can be 
summarized as follows: 

(i) We work in the framework of three generation $SU(2)_L\times U(1)$ model, 
with the addition of three right-handed neutrino fields, which are singlets 
under $SU(2)_L\times U(1)$. No Higgs triplets are introduced and thus the 
effective mass matrix for the left-handed Majorana neutrinos is entirely 
generated by the seesaw mechanism, being given by 
\be
m_L = -m_D M_R^{-1} m_D^T\,,
\label{ss} 
\ee
where $m_D$ denotes the neutrino Dirac mass matrix and $M_R$ stands for the 
Majorana mass matrix of right-handed neutrinos.

(ii) We assume that the neutrino Dirac mass matrix $m_D$ has a hierarchical
eigenvalue structure, analogous to the one for the up-type quarks. This is a
GUT-motivated assumption. However, for our arguments, the only important point 
is that the eigenvalues of $m_D$ be hierarchical, their exact values do not 
play an essential r\^{o}le.

(iii) We assume that the charged lepton and neutrino Dirac mass matrices,  
$m_l$ and $m_D$, are ``aligned'' in the sense that in the absence of the
right-handed mass $M_R$, the leptonic mixing would be small, as it is in the 
quark sector. In other words, we assume that the {\it left-handed} rotations 
that diagonalize $m_l$ and $m_D$ are the same or nearly the same. Again, this 
assumption is motivated by GUTs. We therefore consider that the large lepton 
mixing results from the fact that neutrinos acquire their mass through the 
seesaw mechanism.

(iv) We assume that the Dirac and Majorana neutrino mass matrices are 
unrelated. The exact meaning of this assumption will be explained in the
next section. 

We shall investigate whether the seesaw mechanism, constrained by our
set of assumptions, can lead to a phenomenologically viable neutrino mass 
matrix and, if so, whether it can help us to understand some of the salient 
features of the leptonic mixing. 
In particular, it would be interesting to understand 
why the mixing angle $\theta_{23}$ responsible for the atmospheric
$\nu_\mu\leftrightarrow \nu_\tau$ oscillations is large, while the mixing 
angle $\theta_{13}$ which governs the subdominant $\nu_e\leftrightarrow 
\nu_{\mu(\tau)}$ oscillations of atmospheric neutrinos and long baseline 
$\nu_e\leftrightarrow \nu_{\mu(\tau)}$ oscillations is small. Another 
interesting question is whether this constrained seesaw mechanism can help 
us to discriminate among possible neutrino mass hierarchies -- normal 
hierarchy, inverted hierarchy and quasi-degeneracy. 
Furthermore, it would be useful if the seesaw mechanism could provide some 
guidance as to the possible solutions to the solar neutrino problem -- 
large mixing angle MSW (LMA), small mixing angle MSW (SMA) and vacuum 
oscillations (VO) solutions. We shall address these issues within the 
constrained seesaw mechanism described above. 

\section{General framework}

As previously mentioned, we work in the context of the standard three 
generations $SU(2)_L\times U(1)$ model, where the only additional fields are 
the three right-handed neutrinos $\nu_{iR}$ 
\footnote{Only two of the three known experimental indications of nonzero 
neutrino mass (solar neutrino problem \cite{SNP}, atmospheric neutrino data 
\cite{ANA} and the accelerator LSND results \cite{LSND}) can be explained
through neutrino oscillations with just three light neutrino species. As the 
LSND result is the only one that has not yet been independently confirmed,
we choose not to consider it here.}. 
The most general charged lepton and neutrino mass terms can be written as 
\be
{\cal L}_{mass} = (m_l)_{ij}\,\bar{l}_{iL} l_{jR}+(m_D)_{ij}\,\bar{\nu}_{iL}
\nu_{jR} + \frac{1}{2}(M_R)_{ij}\, \nu_{iR}^T C \nu_{jR} + h.c.\,, 
\label{L}
\ee
where $m_l$ and $m_D$ stand for the charged lepton and neutrino Dirac mass 
matrices arising from Yukawa coupling with the Higgs doublet, while $M_R$ 
denotes the Majorana mass matrix of right-handed neutrinos. Since the 
right-handed Majorana mass terms are $SU(2)_L\times U(1)$ invariant, $M_R$ 
is naturally large, not being protected by the low energy gauge symmetry. 
The matrices $m_l$ and $m_D$ are in general arbitrary complex matrices, while 
$M_R$ is a symmetric complex matrix. 
Without loss of generality we may choose a weak basis (WB) where the charged 
lepton mass matrix is diagonal, with real positive eigenvalues. The lepton 
mass matrices can be written as 
\begin{eqnarray}
m_l\, & \equiv \, & 
d_l =  diag (m_e,\;m_\mu, \; m_\tau)\,, \nonumber \\
m_D & = & V_L\, d_\nu \,V_R^\dagger \,, \nonumber \\
M_R & = & U_R \,D \,U_R^T\,, 
\label{term}
\end{eqnarray}
where $d_l$, $d_\nu$ and $D$ are diagonal, real positive matrices while $V_L$, 
$V_R$ and $U_R$ are unitary matrices. In the absence of $M_R$, the leptonic 
mixing matrix $V$ entering in the charged-current weak interactions would be 
given by $V=V_L$. Our assumption (iii) that $m_l$ and $m_D$ are 
``aligned'' in their left-handed rotations 
means that $V_L$ is assumed to be close to the unit matrix, thus implying that 
in the absence of $M_R$ leptonic mixing would be small, in analogy with the 
quark sector. 

The mass terms in Eq. (\ref{term}) are written in a WB, therefore 
the gauge currents are still diagonal. It should be emphasized that one has 
a large freedom to make WB transformations which leave the gauge currents 
diagonal but alter the mass terms. One can use this freedom to choose, e.g., 
a WB basis where both $m_l$ and $M_R$ are diagonal. However, for our 
arguments, it will be more convenient to choose a different $\nu_R$ basis, 
to be specified below.  

The effective mass matrix of the light left-handed neutrinos 
resulting from the seesaw mechanism can then be written as 
\be
m_L = -V_L\,d_\nu\,W_R\,D^{-1}\,W_R^T\,d_\nu\,V_L^T\,,
\label{ss1}
\ee
where $W_R=V_R^\dagger U_R^*$. The physical leptonic mixing among the light
neutrinos which enters in the probabilities of neutrino oscillations is given 
by the matrix $V$ that diagonalizes $m_L$: 
\be
V^T m_L\,V=diag(m_1,\,m_2,\,m_3)\,.
\label{V}
\ee
Here $m_i$ ($i=1,2,3$) are the masses of the light neutrinos. 
We shall disregard possible CP violation effects in the leptonic
sector and assume the neutrino mass matrix to be real. Its eigenvalues 
$m_i$ can be of either sign, depending on the relative CP parities of
neutrinos. The physical neutrino masses are $|m_i|$. 

One of the challenges is how to obtain large mixing in $V$ 
without resorting to fine tuning. We shall show that 
this is possible in the framework of the seesaw 
mechanism, together with the assumptions listed in sec. 1. 

Following our assumption (iii), we shall consider 
that $V_L\approx 1$ in the WB 
where $m_l$ is diagonal. One can then write
\be
m_L = -d_\nu\,(M_R')^{-1}\,d_\nu\,,
\label{mL1}
\ee
where 
\be
(M_R')^{-1}=W_R\, D^{-1} \, W_R^T \, ,
\label{MR1}
\ee
thus fixing the $\nu_R$ basis. 
It is useful to write the explicit form of $m_L$ as 
\be
m_L =
-\left(\begin{array}{ccc}
m_u^2 \,M_{11}^{-1}  & m_u m_c \,M_{12}^{-1}  & m_u m_t \,M_{13}^{-1} \\
m_u m_c \,M_{12}^{-1}  & m_c^2 \,M_{22}^{-1}  & m_c m_t \,M_{23}^{-1} \\
m_u m_t \,M_{13}^{-1}  & m_c m_t \,M_{23}^{-1}  & m_t^2 \,M_{33}^{-1}
\end{array}
\right)\,,
\label{matr1}
\ee
where $M_{ij}^{-1}\equiv (M_R')^{-1}_{ij}$, and $m_u$, $m_c$ and $m_t$ denote 
the eigenvalues of $m_D$. For our numerical estimates we shall take them
to be equal to the masses of the corresponding up-type quarks, but for our 
general arguments their precise values are unimportant. 

We shall adopt the parametrization of the leptonic mixing matrix $V$ 
which coincides with the standard parametrization of the quark
mixing matrix \cite{PDG} and identify the mixing angle responsible 
for the dominant channel of the atmospheric neutrino oscillations with
$\theta_{23}$, the one governing the solar neutrino oscillations with 
$\theta_{12}$ and the mixing angle which governs the subdominant
$\nu_e\leftrightarrow \nu_{\mu(\tau)}$ oscillations of atmospheric neutrinos 
and long baseline $\nu_e\leftrightarrow \nu_{\mu(\tau)}$ oscillations with 
$\theta_{13}$.
Assuming that $m_1,m_2\ll m_3$ and $\theta_{23}\simeq 45^\circ$ (which is
the best fit value of the Super-Kamiokande data \cite{Ringberg}) 
and taking into account that 
the CHOOZ experiment indicates that $\theta_{13}\ll 1$ \cite{CHOOZ}, it
can be shown that $m_L$ must have the approximate form \cite{Akh}  
\be
m_L = m_0 \left(\begin{array}{ccc}
\kappa      & \varepsilon     & \varepsilon' \\
\varepsilon & ~1+\delta-\delta' & 1-\delta \\   
\varepsilon' & ~1-\delta & 1+\delta+\delta'
\end{array}
\right)\,,
\label{mL2}
\ee
where $\kappa$, $\varepsilon$, $\varepsilon'$, $\delta$ and $\delta'$ are 
small dimensionless parameters. Comparing Eqs. (\ref{matr1}) and (\ref{mL2}), 
one concludes that the following relations should hold, in leading order:
\be
m_c^2\, M_{22}^{-1}=m_t^2\, M_{33}^{-1}=m_c m_t \,M_{23}^{-1}\,.
\label{rel}
\ee
These relations seem to indicate that in order to obtain the form of 
Eq. (\ref{mL2}), strong correlations are required between the entries of 
$d_\nu$ and those of $(M_R')^{-1}$, in apparent contradiction with our 
assumption (iv). However, it can be readily seen that in fact there is no 
conflict. Obviously, the form of $(M_R')^{-1}$ depends on the $\nu_R$ basis 
one chooses. In the definition of $(M_R')^{-1}$ given by Eq. (\ref{MR1}), we 
have included part of the right-handed rotation arising from the 
diagonalization of $m_D$, namely $V_R$ enters in $W_R$ defined as 
$W_R=V_R^\dagger U_R^*$. Therefore $(M_R')^{-1}$ contains information about 
the Dirac mass sector, and Eq. (\ref{rel}) is not necessarily in conflict 
with our assumption (iv).  This assumption has to be formulated in terms
of weak-basis invariants. What should be required is that the ratios of 
the {\it eigenvalues} of $(M_R')^{-1}$ should not be related 
to the ratios of the {\it eigenvalues} of $m_D$. 

In order to see how the phenomenologically favoured 
form of $m_L$ can be achieved without contrived fine tuning between the 
parameters of the Dirac and Majorana sectors, let us first consider the 
two-dimensional sector of $m_L$ in the 2-3 subspace, which is responsible
for a large $\theta_{23}$. We shall write the diagonalized Dirac mass matrix
$d_\nu$ using the dimensionless parameters $p$ and $q$:
\be
d_\nu=m_t\,diag(p^2 q\,,\; p\,,\; 1)\,,\quad p=m_c/m_t\sim 10^{-2}\,, 
\quad q=m_u m_t/m_c^2\sim 0.4\,.
\label{mD}
\ee
It follows from Eq. (\ref{mL1}) that the 2-3 sector of $(M_R')^{-1}$, in
order to lead to the 2-3 structure of Eq. (\ref{mL2}) (with all elements 
approximately equal to unity up to a common factor), should have the
following form: 
\be
M_R^{-1} \propto \left(\begin{array}{cc}
1     & p    \\
p      & p^2   
\end{array}
\right)\,.
\label{M}
\ee
The eigenvalues of the matrix in Eq. (\ref{M}) are 0 and $1+p^2$, and thus by 
choosing the pre-factor to be $const/(1+p^2)$ one arrives at the matrix 
$M_R^{-1}$ of the desired form with $p$- and $q$-independent eigenvalues. 
The question is now whether it is possible to find a $3\times3$ matrix whose 
2-3 sector corresponds to Eq. (\ref{M}) while its eigenvalues (or invariants) 
are independent of $p$ and $q$. This turns out to be possible, and the 
simplest form of $(M_R')^{-1}$ fulfilling this requirement is 
\be
(M_R')^{-1} \propto \frac{1}{1+p^2}
 \left(\begin{array}{ccc}
(1+p^2)\,\gamma & \sqrt{1+p^2}\,(\beta-\alpha p) 
& \sqrt{1+p^2}\,(\alpha+\beta p)\\
\sqrt{1+p^2} \,(\beta-\alpha p) & 1 & p \\   
\sqrt{1+p^2}\,(\alpha+\beta p) & p & p^2
\end{array}
\right)\,,
\label{MR2}
\ee
where the dimensionless parameters $\alpha$, $\beta$ and $\gamma$ do not 
depend on $p$ and $q$. It is straightforward to check that the eigenvalues of 
the matrix in Eq. (\ref{MR2}) are $p$- and $q$-independent. 
The easiest way to do 
that is by noting that $(M_R')^{-1}$ can be written as 
\be
(M_R')^{-1}=S_R^T\,(M_R^0)^{-1} \, S_R
\label{MR3}
\ee
with 
\be
S_R = \left(\begin{array}{ccc}
1   & 0       &  0 \\
0   &  c_\phi & s_\phi \\   
0   & -s_\phi & c_\phi
\end{array}
\right)\,,\quad 
(M_R^0)^{-1} = \frac{1}{2 M}\left(\begin{array}{ccc}
\gamma   & \beta  & \alpha \\
\beta    & 1      & 0   \\   
\alpha   & 0      & 0
\end{array}
\right)\,, 
\label{SR}
\ee
and 
\be
c_\phi=\cos \phi\,,\quad\quad s_\phi=\sin\phi\,,\quad\quad \phi=\arctan p\,.
\label{phi}
\ee
{}From Eqs. (\ref{mL1}), (\ref{MR3}), (\ref{SR}) and (\ref{phi}) one obtains
\be
m_L = \frac{m_t^2}{2M}\frac{p^2}{1+p^2}\left(\begin{array}{ccc}
{q'}^2 p^2\,\gamma  & q' p\,(\beta-\alpha p) & q'\,(\alpha+\beta p) \\
q' p\, (\beta-\alpha p) & 1 & 1  \\   
q'\, (\alpha+\beta p) & 1 & 1
\end{array}
\right)\,,
\label{mL3}
\ee
where 
\be
q' \equiv q \sqrt{1+p^2} \simeq q\,.
\label{q}
\ee
It is worth emphasizing that we have obtained $m_L$ of the desired form, 
while abiding by our assumptions. 
Comparison of Eqs. (\ref{mL2}) and (\ref{mL3}) leads to the following 
identification for the parameters of the phenomenological mass matrix of 
light neutrinos:
\be
\kappa = {q'}^2 p^2\,\gamma\,, \quad\quad \varepsilon = q' p\,(\beta-\alpha p) 
\,,\quad\quad \varepsilon'= q'\,(\alpha+\beta p)\,, \quad \delta=\delta'=0\,.
\label{param1}
\ee
The largest eigenvalue of the matrix $m_L$ in Eq.
(\ref{mL3}), 
i.e. the mass of 
the heaviest of the three light neutrinos is 
\be
m_3 \simeq \frac{m_t^2}{M} \frac{p^2}{1+p^2}\simeq \frac{m_c^2}{M}\,. 
\label{m3}
\ee
It scales as $m_c^2$ rather than as usually expected $m_t^2$. It has to be 
identified with $\Delta m^2_{atm} \simeq (2 - 6)\times 10^{-3}$ eV$^2$, which 
gives 
\be
M\simeq (10^{10} - 10^{11})~{\rm GeV}\,,
\label{MM}
\ee
i.e. an intermediate mass scale rather than the GUT scale. 

It has been shown in \cite{Akh} that the MSW effect \cite{MSW} can only occur 
for neutrinos, and in particular the LMA and SMA solutions of the solar
neutrino problem are only possible, if the parameters of the mass matrix
$m_L$ in Eq. (\ref{mL2}) satisfy 
\be
| 4\delta-\delta'^{2}|> | 2{\kappa}-(\varepsilon^2+\varepsilon'^{2})|\,.
\label{cond2}   
\ee
Since $\delta=\delta'=0$ in Eq. (\ref{mL3}), it is clear that
the only solution to the solar neutrino problem that is not
automatically ruled out in the case under consideration
is the VO solution. 

It is interesting to ask whether our scheme can be modified so that 
nonzero values for $\delta$ and $\delta'$ be obtained. One simple 
possibility would be to assume an incomplete alignment between the mass 
matrix of charged leptons and the Dirac mass matrix of neutrinos:
$V_L\approx 1$ instead of $V_L=1$. Then the effective mass matrix of light 
neutrinos $m_L$ would be obtained from Eq. (\ref{mL3}) by the additional 
rotation by $V_L$. Taking for simplicity this rotation to be in the 2-3 
subspace, one can readily make sure that it indeed yields nonzero $\delta$ 
and $\delta'$. However, in this case they are related by $\delta=\delta'^2/4$. 
Therefore the left-hand side (l.h.s.) of (\ref{cond2}) vanishes, i.e. this 
inequality is not satisfied and the MSW solutions of the solar neutrino 
problem are still not possible. The fact that an additional rotation in
the 2-3 subspace does not change the l.h.s. of  (\ref{cond2}) becomes obvious 
by noticing that $4\delta-\delta'^2$ coincides with the determinant of 
the $2\times 2$ submatrix of Eq. (\ref{mL2}) in the 2-3 subspace. 
Thus, to accommodate the LMA or SMA solutions of the solar neutrino problem 
through the $V_L$ rotation one should consider a matrix $V_L$ of a more 
general form. 

There is, however, another way to achieve the same goal. One can arrive at 
$\delta, \delta' \neq 0$ even with $V_L=1$ if one considers the following 
simple modification of the the matrix $(M_R^0)^{-1}$ in Eq. (\ref{SR}):
\be
(M_R^0)^{-1} = \frac{1}{2 M}\left(\begin{array}{ccc}
\gamma   & \beta  & \alpha \\
\beta    & 1      & 0   \\   
\alpha   & 0      & r
\end{array}
\right)\,, 
\label{MR0}
\ee
i.e. the 33-element of the matrix now is nonzero. The requirement 
$|\delta|, |\delta'| \ll 1$ translates into $|r|\ll p^2$. 
This yields the following effective mass matrix for the light neutrinos: 
\be
m_L \simeq \frac{m_t^2}{2M}\frac{p^2}{1+p^2}\left(\begin{array}{ccc}
{q'}^2 p^2\,\gamma  & q' p\,(\beta-\alpha p) & q'\,(\alpha+\beta p) \\
q' p\, (\beta-\alpha p) & 1-r/4p^2 & 1-r/4p^2  \\   
q'\, (\alpha+\beta p) & 1-r/4p^2 & 1+3r/4p^2
\end{array}
\right)\,,
\label{mL4}
\ee
This means that now 
\be
\delta \simeq r/4p^2\,, \quad\quad \delta' \simeq r/2p^2\,,
\label{deltas}
\ee
and so the l.h.s. of (\ref{cond2}) is nonzero, i.e. SMA and LMA solutions 
of the solar neutrino problem are possible. The parameters $\kappa$, 
$\varepsilon$ and $\varepsilon'$ in this case are the same as in the 
case $r=0$, i.e. are given by Eq. (\ref{param1})
\footnote{The particular case of the neutrino mass matrix of the form
(\ref{mL4}) with $\beta=\gamma=r=0$ (which allows only the VO solution 
of the solar neutrino problem) was obtained in \cite{JS}.}. 

\section{Numerical examples}

We shall now give some illustrative examples of the values of the 
parameters for which all three types of the solutions -- SMA, LMA and 
VO -- are possible. We concentrate here on the case of normal mass 
hierarchy ($m_{1}, m_{2}\ll m_3$); the cases of inverted mass hierarchy 
and quasi-degeneracy will be discussed in sec. 4.

The neutrino mass squared differences which enter in the probabilities of the 
solar and atmospheric neutrino oscillations are related to the eigenvalues 
of the effective mass matrix of light neutrinos $m_L$ via $\Delta m_\odot^2 
\simeq \Delta m_{21}^2$, $\Delta m_{atm}^2\simeq \Delta m_{31}^2 \simeq m_3^2$. 
Consider first the following choice of the parameters of the matrix 
$(M_R^0)^{-1}$ in Eq. (\ref{MR0}): $\alpha= 1.1\cdot 10^{-2}$, $\beta\sim
\alpha$, $\gamma \aprle \beta^2$, $r=8\cdot 10^{-6}$. This gives the
following values of the parameters of the mass matrix of light neutrinos 
$m_L$ in (\ref{mL2}): $\kappa \aprle 10^{-9}$, $\varepsilon\simeq 3\cdot 
10^{-5}$, $\varepsilon'\simeq 4.5\cdot 10^{-3}$, $\delta \simeq 0.04$,
$\delta' \simeq 0.08$. 
Diagonalization of $m_L$ can then be easily performed either by making use 
of the approximate analytic expressions derived in \cite{Akh} or 
numerically. It gives the following values of the masses and mixings of light 
neutrinos:
\[
m_1\simeq -3.6\cdot 10^{-6} {\rm eV}\,,\quad
m_2\simeq 2.3\cdot 10^{-3} {\rm eV}\,,\quad
m_3\simeq  2 m_0 \simeq 0.06\; {\rm eV}\,,
\]
\be
\sin^2 2\theta_{12}\simeq 6.2\cdot 10^{-3}\,,\quad 
\sin\theta_{13}\simeq 1.7\cdot 10^{-3}\,,
\label{SMA}
\ee
where we have taken $m_0\simeq \sqrt{\Delta m_{atm}^2}\,/2\simeq 0.03$ eV.
In all the cases we consider, the value of the mixing angle $\theta_{23}$
is very close to $45^\circ$ by construction of our mass matrix $m_L$. 
{}From Eq. (\ref{SMA}) one finds $\Delta m_\odot^2\simeq 5.3\cdot 10^{-6}$ 
eV$^2$, i.e. this choice of the parameters leads to the SMA solution of
the solar neutrino problem
\footnote{For recent analyses of the solar neutrino data and allowed 
ranges of the parameters see \cite{BKS1,Valencia,BKS2}.}. 
The corresponding mass eigenvalues of heavy Majorana neutrinos are 
$M_1\simeq 6\cdot 10^{10}$ GeV, $M_2\simeq -M_3\simeq 5.5\cdot 10^{11}$ GeV. 

Let us now choose $\alpha= -0.75$, $\beta\sim \alpha$, $\gamma \aprle
\beta^2$, $r=1.5\cdot 10^{-5}$. This yields $\kappa \aprle 4\cdot 10^{-6}$, 
$\varepsilon\simeq 2\cdot 10^{-3}$, $\varepsilon'\simeq 0.3$, $\delta
\simeq 0.073$, $\delta' \simeq 0.146$. Diagonalization of $m_L$ then gives 
\[
m_1\simeq -4.62\cdot 10^{-3} {\rm eV}\,,\quad
m_2\simeq 7.86\cdot 10^{-3} {\rm eV}\,,\quad
m_3\simeq  2 m_0 \simeq 0.06\; {\rm eV}\,,
\]
\be
\sin^2 2\theta_{12}\simeq 0.83\,, \quad 
\sin\theta_{13}\simeq 0.11\,,
\label{LMA}
\ee
with $\Delta m_\odot^2\simeq 4\cdot 10^{-5}$ eV$^2$, i.e. this choice of
the parameters leads to the LMA solution of the solar neutrino problem. 
The corresponding mass eigenvalues of heavy Majorana neutrinos are 
$M_1\simeq 4.6\cdot 10^{10}$ GeV, $M_2\simeq -7.4\cdot 10^{10}$ GeV, 
$M_3 \simeq 1.1\cdot 10^{11}$ GeV. 

Finally, let us choose $\alpha=1.13\cdot 10^{-3}$, $\beta\aprle \alpha$,
$\gamma \aprle \beta^2$, $r=2\cdot 10^{-8}$. This yields $\kappa \aprle
10^{-11}$, $\varepsilon\aprle 3\cdot 10^{-6}$, $\varepsilon'\simeq 4.5\cdot 
10^{-4}$, $\delta \simeq 10^{-4}$, $\delta' \simeq 2\cdot 10^{-4}$.
Diagonalization of $m_L$ gives 
\[
m_1\simeq -6.95\cdot 10^{-6} {\rm eV}\,,\quad
m_2\simeq 1.29\cdot 10^{-5} {\rm eV}\,,\quad
m_3\simeq  2 m_0 \simeq 0.06\; {\rm eV}\,,
\]
\be
\sin^2 2\theta_{12}\simeq 0.91\,, \quad 
\sin\theta_{13}\simeq 1.6\cdot 10^{-4}\,,
\label{VO}
\ee
with $\Delta m_\odot^2\simeq 1.2\cdot 10^{-10}$ eV$^2$, i.e. this choice
of the parameters leads to the VO solution of the solar neutrino problem. 
The corresponding mass eigenvalues of heavy Majorana neutrinos are 
$M_1\simeq 6\cdot 10^{10}$ GeV, $M_2\simeq -M_3\simeq 5.3\cdot 10^{13}$
GeV. 
Alternatively, one could choose $\alpha=1.75\cdot 10^{-2}$, $\beta\aprle 
\alpha$, $\gamma \aprle \beta^2$, $r=0$. This gives $\kappa \aprle 2.5\cdot
10^{-9}$, $\varepsilon\aprle 5\cdot 10^{-5}$, $\varepsilon'\simeq 7\cdot 
10^{-3}$, $\delta=\delta' =0$. One then obtains $\Delta m_\odot^2 
\simeq 1.1\cdot 10^{-10}$ eV$^2$, $\sin \theta_{13} \simeq 2.5\cdot 10^{-3}$ 
and $\theta_{12}\simeq \theta_{23}\simeq 45^\circ$, i.e. this choice of
the parameters leads to the VO solution of the solar neutrino problem with 
bi-maximal neutrino mixing \cite{bimax,JS}. 
The mass eigenvalues of heavy Majorana neutrinos in this case are
$M_1\simeq 6\cdot 10^{10}$ GeV, $M_2\simeq -M_3\simeq 3.4\cdot 10^{12}$ GeV. 

The examples given here demonstrate that with the inverse mass matrix 
of right-handed neutrinos $(M_R')^{-1}$ of the form (\ref{MR0}), depending 
on the values of its parameters, all three main neutrino oscillations 
solutions to the solar neutrino problem -- SMA, LMA and VO -- can be realized 
in the framework of the constrained seesaw mechanism.

\section{Inverted mass hierarchy and quasi-degeneracy}

We shall now briefly discuss the other possible neutrino mass hierarchies
-- the inverted mass hierarchy $|m_3|\ll |m_1|\simeq |m_2|$ and 
the quasi-degenerate case with $|m_1|\simeq |m_2|\simeq |m_3|$. For the case 
of the inverted mass hierarchy there are three main zeroth order textures
corresponding to the limit $|m_1|=|m_2|$, $\theta_{13}=0$, 
$\theta_{23}=45^\circ$ (see, e.g., \cite{Alt}):  
\be
m_L \propto \left(\begin{array}{ccc}
\pm 2     & 0     & 0 \\
0         & 1     & 1\\   
0         & 1     & 1
\end{array}
\right)\,; \quad
\left(\begin{array}{ccc}
0     & 1     &   1 \\
1     & 0     &   0  \\   
1     & 0     &   0
\end{array}
\right)\,. 
\label{text}
\ee
The first two textures differ only by the sign of the 11-term. One can
then invert Eq. (\ref{mL1}) to find the inverse mass matrices of right-handed 
neutrinos $(M_R')^{-1}$ which lead to these textures. To be consistent
with our assumption (iv), these matrices must satisfy the following 
requirement: it should be possible to obtain each of them by a $p$- and
$q$-dependent rotation of a $p$- and $q$-independent matrix. In other words, 
their eigenvalues $\Lambda_i$ ($i=1,2,3$) must be $p$- and $q$-independent. 

Since all mass matrices in Eq. (\ref{text}) have one zero eigenvalue, so
do the corresponding matrices $(M_R')^{-1}$, and their determinants vanish. 
Therefore it is sufficient to check if their traces ($\Lambda_1+\Lambda_2+
\Lambda_3$) and second invariants ($\Lambda_1\Lambda_2+\Lambda_1\Lambda_3+
\Lambda_2\Lambda_3$) can be made $p$- and $q$-independent by a proper
choice of the pre-factors in Eq. (\ref{text}). 
One can readily make sure that for the first two textures (first matrix in
Eq. (\ref{text}) with both signs of the 11 element) this is impossible: 
if the trace of the corresponding matrix $(M_R')^{-1}$ is made $p$- and 
$q$-independent, then the second invariant depends on $p$ and $q$, and 
vice versa. Therefore these zeroth-order textures do not satisfy our 
conditions (i) -- (iv). Perturbing these textures by adding small terms  
to each their element will not change this conclusion. 

The situation is different for the second matrix in Eq. (\ref{text}). 
The corresponding matrix $(M_R')^{-1}$ has zero determinant and trace, 
and the second invariant can be always made $p$- and $q$-independent by a 
proper choice of the pre-factor. However, this texture is ruled out 
on different grounds. It describes the bi-maximal mixing with the inverted 
hierarchy, which means $\theta_{12}=45^\circ$. In this case only the 
VO solution of the solar neutrino problem is possible \cite{BKS1,Valencia}.  
This implies $\Delta m^2_{21}=\Delta m^2_\odot \sim 10^{-10}$ eV$^2$. 
Small nonvanishing values of $\Delta m^2_{21}$ are 
achieved when zeros in the texture matrix are filled in with small terms, 
and for the VO solution these small terms should be $\aprle 10^{-8}$. 
The diagonalization of the corresponding matrix $(M_R')^{-1}$ then 
gives the following values of the masses of the heavy singlet neutrinos
$M_i=\Lambda_i^{-1}$: two singlet neutrinos are almost degenerate with 
$M_1\simeq -M_2\sim 10^8$ GeV; the third mass eigenvalue turns out to be 
well above the Planck scale:  $M_3\sim 10^{22}$ GeV, clearly not a physical 
value. Thus, this case of the inverted mass hierarchy is ruled out as well. 
It is interesting to note that in our argument we have not used the condition 
(iv), i.e. the mass matrices obtained from the last texture in 
Eq. (\ref{text}) are excluded on the basis of our assumptions (i)-(iii) only. 
The mass matrices leading to the quasi-degenerate neutrino mass spectrum can 
be considered and ruled out using arguments analogous to those applied to 
the cases of the first two textures in Eq. (\ref{text}).

\section{Discussion}

We have shown that the seesaw mechanism, supplemented by the set of  
assumptions listed in the Introduction, leads to phenomenologically viable
mass matrices of light active neutrinos. The mixing angle $\theta_{23}$ 
responsible for the dominant channel of the atmospheric neutrino oscillations 
can be naturally large without any fine tuning. The fact that the seesaw
mechanism can lead to large lepton mixing even if the mixing in both the
Dirac and Majorana sectors is small has been known for some time  
\cite{AFM,Sm,Tan}. Our approach gives a simple explanation to this fact: if 
the eigenvalues of the Dirac mass matrix of neutrinos $m_D$ are hierarchical, 
and the entries $M_{ij}^{-1}$ of the inverse Majorana mass matrix 
$(M_R')^{-1}$ have the hierarchy $M_{22}^{-1}\gg M_{23}^{-1} \gg M_{33}^{-1}$ 
(which implies small mixing in the 2-3 sector of right-handed Majorana 
neutrinos), then the multiplication of $(M_R')^{-1}$ by $m_D$ from the left 
and from the right suppresses the 22 element of the resulting effective mass 
matrix $m_L$ to a larger extent than it suppresses the 23 element, which in 
turn is more suppressed than the 33 element. This can lead to all the elements 
of the 2-3 sector of the resulting matrix $m_L$ being of the same order,
yielding a large mixing angle $\theta_{23}$. 

Although the constrained seesaw mechanism allows to obtain a large mixing
angle $\theta_{23}$ in a very natural way, it does not explain why 
$\theta_{23}$ is large: the largeness of this mixing angle is merely related 
to the choice of the inverse mass matrix of heavy singlet neutrinos, Eq.  
(\ref{SR}) or (\ref{MR0}). However, once this choice has been made, the 
smallness of the mixing angle $\theta_{13}$ which determines the element 
$V_{e3}$ of the lepton mixing matrix can be readily understood. For the
case of the normal mass hierarchy $m_1, m_2 \ll m_3$ the value of 
$\theta_{13}$ can be expressed in terms of the entries of the effective mass 
matrix $m_L$ in Eq. (\ref{mL2}) as $\sin\theta_{13}\simeq(\varepsilon+ 
\varepsilon')/2\sqrt{2}$ \cite{Akh}. From Eq. (\ref{param1}) one then finds   
$\sin\theta_{13}\simeq q(\alpha+2\beta p)/2\sqrt{2} \simeq q\alpha/2\sqrt{2}
\simeq 0.14 \alpha$, assuming $\beta \ll \alpha p^{-1}\sim 100 \alpha$.
Since all the solutions of the solar neutrino problem require $|\alpha|<1$ 
in order to have small enough $\Delta m_\odot^2$ (see sec. 3), the smallness 
of $\theta_{13}$ follows. 
 
We have shown that all three main neutrino oscillations solutions to the
solar neutrino problem -- small mixing angle MSW, large mixing angle MSW 
and vacuum oscillations -- are possible within the constrained seesaw.  
The mechanism does not favour any of these solutions over the others. 

The seesaw mechanism we have studied naturally leads to the normal neutrino 
mass hierarchy while disfavouring the inverted mass hierarchy and 
quasi-degenerate neutrinos. 
For LMA and SMA solutions of the solar neutrino problem, the masses of the 
heavy singlet neutrinos are of the order $10^{10} - 10^{11}$ GeV. For 
the VO solution, the lightest of the singlet neutrinos has the mass 
of the same order of magnitude, whereas the masses of the other two 
are $\sim 10^{12} - 10^{13}$ GeV.  

\vglue 0.3truecm
The authors are grateful to K.S. Babu, M. Lindner, P. Lipari, M. Lusignoli, 
R.N. Mohapatra, G. Senjanovi\'c and A.Yu. Smirnov for useful discussions.
This work was supported in part by the TMR network grant ERBFMRX-CT960090
of the European Union. The work of E.A. was supported by Funda\c{c}\~ao
para a Ci\^encia e a Tecnologia through the grant PRAXIS XXI/BCC/16414/98.

\end{document}